\documentclass[preprint2,10.5pt]{aastex}
\begin{document}
\title{\large{\rm{ON THE METALLICITY DEPENDENCE OF CLASSICAL CEPHEID LIGHT AMPLITUDES}}}
\author{D. Majaess$^1$, D. G. Turner$^2$, W. Gieren$^3$, L. Berdnikov$^{4,5}$, D. J. Lane$^2$}
\affil{$^1${\rm \small Halifax, Nova Scotia, Canada.}}
\affil{$^2${\rm \small Saint Mary's University, Halifax, Nova Scotia, Canada.}}
\affil{$^3${\rm \small Universidad de Concepci\'on, Concepci\'on, chile.}}
\affil{$^4${\rm \small Moscow M V Lomonosov State University, Sternberg Astronomical Institute, Moscow, Russia.}}
\affil{$^5${\rm \small Isaac Newton Institute of Chile, Moscow branch, Moscow, Russia.}}
\email{\rm dmajaess@cygnus.smu.ca}

\begin{abstract}
Classical Cepheids remain a cornerstone of the cosmic distance scale, and thus characterizing the dependence of their light amplitude on metallicity is important.  Period-amplitude diagrams constructed for longer-period classical Cepheids in IC 1613, NGC 3109, SMC, NGC 6822, LMC, and the Milky Way imply that very metal-poor Cepheids typically exhibit smaller $V$-band amplitudes than their metal-rich counterparts. The results provide an alternate interpretation relative to arguments for a null and converse metallicity dependence. The empirical results can be employed to check predictions from theoretical models, to approximate mean abundances for target populations hosting numerous long-period Cepheids, and to facilitate the identification of potentially blended or peculiar objects. 
\end{abstract}
\keywords{stars: oscillations, stars: abundances, stars: variables: Cepheids}

\section{{\rm \footnotesize INTRODUCTION}}
Establishing an empirically derived correlation between a classical Cepheid's light amplitude, pulsation period, and chemical composition is important for several reasons. For example, such a pertinent link is needed to help constrain theoretical models.  Model-generated period-amplitude diagrams are acutely sensitive to the input physics, the adopted mass-luminosity relation, the inclusion of convective overshooting \citep[][their Fig.~6]{bo00} and rotation, the spatial resolution of the ionization zones \citep[][their Fig.~10]{pe03}, and the generated amplitudes once transformed to Johnson $V$ can be $\sim20$\% too large \citep[][their Fig.~7]{bo00}.  \citet{bo00} found that metal-poor Cepheids typically exhibit larger amplitudes than their metal-rich counterparts, except for the 7 $M_{\sun}$ canonical and 9 $M_{\sun}$ convective overshooting models.  Thus metal-rich $10-30^{\rm d}$ ($7-9M_{\sun}$) Cepheids may potentially exhibit larger amplitudes than their metal-poor counterparts.  The availability of such testable model predictions is desirable, however, the diversity of opinion concerning the empirical picture (a partial account is provided below) hindered a comparison.

\begin{figure*}[!t]
\begin{center}
\includegraphics[width=18.6cm]{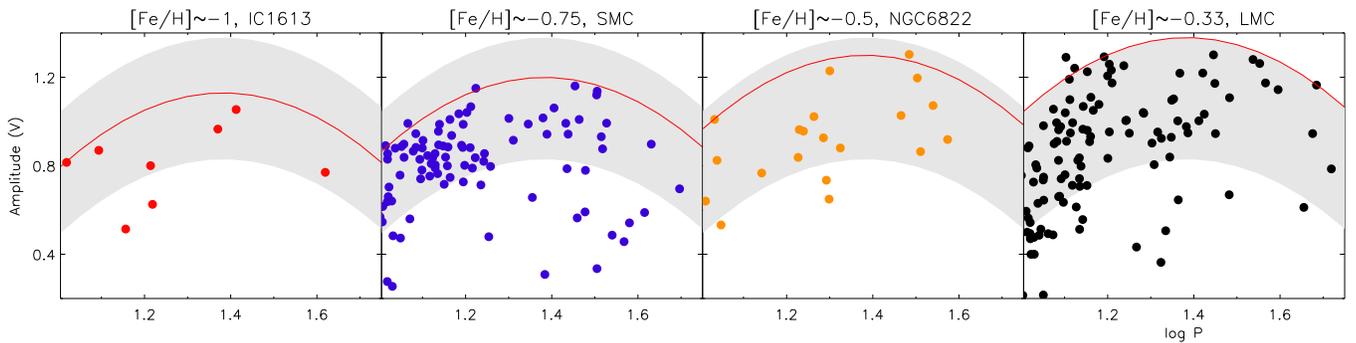} 
\caption{\small{Period-amplitude diagrams for classical Cepheids in IC 1613 (red, $[{\rm Fe/H}]\sim-1$), SMC (purple, $[{\rm Fe/H}]\sim-0.75$), NGC 6822 (orange, $[{\rm Fe/H}]\sim-0.5$), and the LMC (black, $[{\rm Fe/H}]\sim-0.33$). The galaxies are listed in order of increasing metal abundance (the shaded gray region characterizes Milky Way Cepheids, $[{\rm Fe/H}]\sim0$).  The results imply that longer-period metal-rich classical Cepheids (e.g., Milky Way) typically exhibit larger $V$-band amplitudes than their very metal-poor counterparts (e.g., SMC). Overlaid red lines trace an approximate mean upper bound to the distribution, which appears to increase in tandem with metallicity up to the LMC.  Cepheids deviating from the lower-envelope delineated by the bulk distribution may suffer from blending (binarity or crowding), or may be peculiar.}}
\label{fig-pa}
\end{center}
\end{figure*}

Fig.~1 in \citet{sh42} highlights the sizable amplitude offset between $\sim300+$ SMC and Galactic Cepheids.  \citet{sh42} argued that the discrepancy may be tied to systematic errors, whereas \citet{ak61} suggested the offset was attributable to differences in chemical composition between SMC and Galactic Cepheids ($\Delta [{\rm Fe/H}]\sim-0.75$, \citealt{lu98,ro08}).  \citet{st70} instead noted a general similarity among period-amplitude relations characterizing Cepheids in the Galaxy and Magellanic Clouds.  \citet[][and references therein]{vg78} argued otherwise, namely that longer-period metal-poor classical Cepheids exhibit smaller amplitudes than their metal-rich counterparts, thereby supporting the results of \citet{ak61}.   \citet{bi86} and \citet{pp00} corroborated the \citet{vg78} conclusion concerning the amplitude-metallicity behavior of longer-period Cepheids.  A recent analysis by \citet{sk12} indicated that metal-poor classical Cepheids display slightly larger amplitudes than their metal-rich counterparts, a result inferred by correlating the latest abundance and photometric data for Galactic Cepheids ($-0.4\lesssim[{\rm Fe/H}]\lesssim0.4$).  \citet{sk12} found the correlation strongest for shorter-period Cepheids, whereas the impact of metallicity on radial velocity and $B$-band amplitudes for longer-period Cepheids was less pronounced, and the coefficient for the latter was not significant (their Table 2 and Fig.~4).  

At present the template lightcurve repository used to analyze sparse extragalactic Cepheid observations features a mix of SMC, LMC, and Galactic Cepheids \citep[][]{st96,ta05,yo09}, without regard for metallicity spread among the calibrating and parent galaxies. \citet{ta05} and \citet{yo09}\footnote{The \citet{yo09} code for fitting template lightcurves to sparse Cepheid observations is available @ \url{http://www.astro.washington.edu/users/yoachim/code.php}} hinted that an amplitude offset between the samples could be linked to metallicity, whereas \citet{st96} suggested that a minor offset might be exaggerated in part by selection effects tied to Galactic Cepheids \citep[see also][]{sh42,sz03a}, and indeed, results presented in \S \ref{s-analysis} confirm that an offset is difficult to discern between relatively metal-rich Galactic and LMC Cepheids.  Extending the comparison to include very metal-poor galaxies such as the SMC and IC 1613 is thus desirable, since the amplitude offset presumably increases in concert with the metallicity baseline established.  Ultimately, constraining the amplitude-metallicity relation may help mitigate uncertainties associated with fitting lightcurve templates to sparsely observed (HST) Cepheids.

In this study period-amplitude diagrams are constructed for longer-period classical Cepheids in the Milky Way ($[{\rm Fe/H}]\sim0$), LMC ($[{\rm Fe/H}]\sim-0.33$), NGC 6822 ($[{\rm Fe/H}]\sim-0.5$), SMC ($[{\rm Fe/H}]\sim-0.75$), and IC 1613 ($[{\rm Fe/H}]\sim-1$).  The galaxies span a sizable metallicity baseline ($\Delta [{\rm Fe/H}]\sim-1$), and the latest observational data for the comparatively uncrowded galaxies are analyzed (e.g., OGLE III).  Longer-period very metal-poor classical Cepheids are found to typically exhibit smaller $V$-band amplitudes than their metal-rich counterparts. The difference is sufficiently discernible that the period-amplitude diagram may be employed in certain cases to approximate the mean metallicity for a target population containing numerous long-period classical Cepheids, provided the sample is relatively free from the effects of blending. The results may be employed to support metallicity estimates inferred from adjacent young stars and nebulae (e.g., NGC 3109, \S \ref{s-ngc3109}).

\begin{figure}[!t]
\begin{center}
\includegraphics[width=9cm]{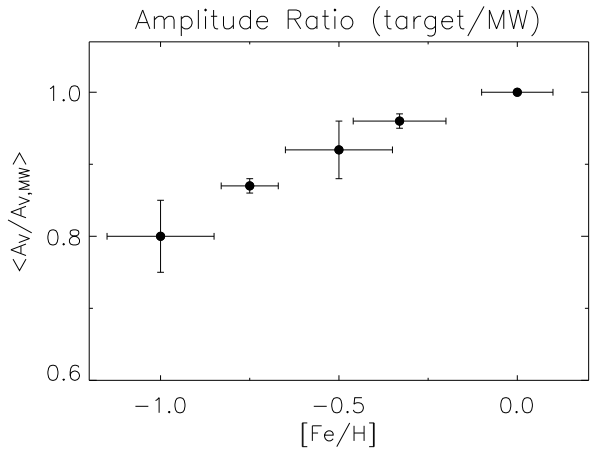} 
\caption{\small{To assess the impact of the differing sample sizes on the perceived amplitude offset, Cepheids were randomly drawn from the Milky Way (MW) dataset until the total typically equalled that of the comparison dataset (e.g., IC 1613, $[{\rm Fe/H}]\sim-1$).  Median amplitude ratios were subsequently inferred from 50+ simulations, and were evaluated relative to MW Cepheids.  The uncertainties displayed are the standard deviations ($\sigma [{\rm Fe/H}]=0.15$ dex was assumed for IC 1613 and NGC 6822).  Longer-period metal-rich Cepheids typically exhibit larger $V$-band amplitudes than their very metal-poor counterparts.}}
\label{fig-med}
\end{center}
\end{figure}

\begin{figure}[!t]
\begin{center}
\includegraphics[width=8cm]{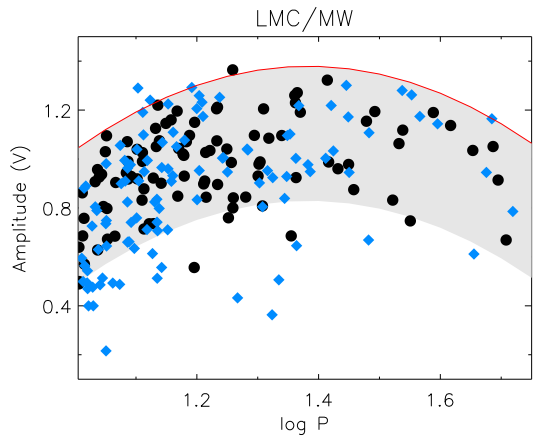} 
\caption{\small{Period-amplitude diagram for classical Cepheids in the Milky Way (black dots, $[{\rm Fe/H}]\sim0\pm0.1(\sigma)$) and LMC (blue diamonds, $[{\rm Fe/H}]\sim-0.33\pm0.13(\sigma)$).  Differences between the samples are difficult to discern, which is likely attributable in part to overlap (LMC Cepheids span $-0.62\leq[{\rm Fe/H}]\leq-0.1$, \citealt{mo06}).}}
\label{fig-pa3}
\end{center}
\end{figure}

\section{{\rm \footnotesize ANALYSIS}}
\label{s-analysis}
\subsection{{\rm \footnotesize PHOTOMETRY}}
\label{s-phot}
Only longer-period classical Cepheids ($P>10^{\rm d}$) are examined here, since for surveys of remote galaxies such stars typically exhibit smaller photometric uncertainties owing to their intrinsic brightness. Uncertainties introduced by misclassified short-period overtone pulsators and undersampled fainter objects are eliminated. Long-period classical Cepheids also appear less sensitive to blending \citep[e.g.,][their Fig.~17]{ma06}. A Cepheid's amplitude can be artificially reduced as a result of blending, so mitigating that effect is important.  

Photometric parameters for Galactic Cepheids were derived from the expansive compilation of \citet{be08} \citep[see also][]{be00}, and supplemented in certain instances using observations acquired from the Abbey-Ridge Observatory \citep[e.g., V396 Cyg, S Vul, etc.,][]{la08}.  Photometry for variables in IC 1613 and the Magellanic Clouds was taken from the OGLE survey \citep[e.g.,][]{ud01,so08,so10}. Araucaria photometry for Cepheids in NGC 6822 was published by \citet{pi04}. Following \citet{ks09} and \citet{sk12}, peak-to-peak amplitudes were determined for the Cepheids \citep[see also the pertinent discussion by][]{ng03}. The period-amplitude results derived for the Galactic sample are similar to those found in the comprehensive analysis by \citet{ks09}.  $V$-band observations were selected as reliable $B$-band extragalactic data are sparse, and the amplitude diminishes with increasing wavelength \citep[][their Fig.~5]{mf91}.  Johnson $V$ is also common and well standardized, thus reducing uncertainties introduced by mismatched filters (e.g., $U$-band).

\subsection{{\rm \footnotesize AMPLITUDE-METALLICITY DEPENDENCE}}
A comparison between classical Cepheids in the Galaxy and IC 1613 is particularly pertinent granted the abundance difference between the samples is sizable, and approximately $\Delta [\rm{Fe/H}]\sim-1$ \citep{lu98,ta07,br07}. Fig.~\ref{fig-pa} demonstrates that metal-poor Cepheids in IC 1613 typically have smaller $V$-band amplitudes than their Galactic counterparts, although a larger sample is desirable.  However, a simulation featuring $50+$ samplings from the Milky Way data, such that the total number of Cepheids randomly drawn equals the sample size of IC 1613, yields a median amplitude ratio of: $0.80\pm 0.05 (\sigma )$ (IC 1613/MW, Fig.~\ref{fig-med}). That evaluation suggests, in concert with the results for the SMC, that the amplitude offset is probably not attributable to poor statistics.  The median amplitude ratio can be \textit{approximated} as:
\begin{eqnarray}
\label{eqn-m}
\nonumber
<A_V/A_{V,MW}>\sim -0.1{\rm[Fe/H]}^2 +0.1{\rm[Fe/H]}+1 \\ 
(10\leq P \leq 30^{\rm d}, -1 \leq {\rm[Fe/H]} \leq 0) 
\end{eqnarray}
That equation will require updating as new data become available.  Figs.~\ref{fig-pa} and \ref{fig-med} may be used in unison to gauge the abundance of a target population, namely so to mitigate the impact of high and low-amplitude outliers.  Cepheids in the SMC and NGC 6822 are likewise more metal-poor than Galactic Cepheids, and the amplitudes of Cepheids in the SMC are noticeably smaller than those of their Galactic counterparts. Similar to IC 1613, a comparison between Galactic and SMC classical Cepheids is pertinent granted the abundance difference between the two samples is large, $\Delta [\rm{Fe/H}]\sim-0.75$ \citep{mo06}.  Differences are not readily apparent between Milky Way and LMC Cepheids (Figs.~\ref{fig-med}, \ref{fig-pa3}), a fact which is explained in part by the metallicity overlap between those populations \citep{mo06,lu11}.  Figs.~\ref{fig-med} and \ref{fig-pa3} may be consistent in part with findings inferred from \citet{sk12}, and indicate that amplitudes for longer-period Cepheids in the $[{\rm Fe/H}]\gtrsim-0.4$ regime may display a marginal dependence on metallicity (note that Fig.~\ref{fig-med} is described by a converging polynomial: Eqn.~\ref{eqn-m}). 

It is desirable to extend the analysis of \citet{lu98} and \citet{mo06} to establish direct abundances for Cepheids in NGC 6822 and IC 1613. Abundance estimates by those authors were inferred from Cepheid spectra, which consequently obviates concerns regarding whether abundances deduced from nebulae (or otherwise) are representative of the Cepheid population \citep[see also][]{br09,br10}.   For example, the abundance estimate for NGC 6822 stems from two A-type supergiants in that galaxy (\citealt{ve01}, $[\rm{Fe/H}]=-0.49\pm0.20$).   Abundances inferred from nebulae seem to confirm the \citet{ve01} result that young stars in NGC 6822 are more metal-rich than their SMC counterparts.  Yet nebular abundances are also uncertain, as evidenced by the debate concerning the galactocentric metallicity gradients in M33 and M106 \citep{bo08,ma10,br11a,br11b}.  

The physical basis for the impact of metallicity on Cepheid light amplitudes is complicated \citep[see also][and discussion therein]{st70,sk12}.  It is unclear whether very metal-poor longer-period Cepheids are preferentially sampling the red-edge of the instability strip, where the pulsation is damped owing to convection.  In older stellar evolutionary models increased metallicity depressed high surface temperature excursions for lower-mass ($<5 M_{\sun}$) stars during the blue-loop stage following core helium ignition \citep{be85}.  However, the extent of blue-loops may likewise depend on CNO abundances in the envelopes \citep{xl04}, which are governed in part by rotation and meridional mixing during the main-sequence stage \citep{ma01,tb04}.  The empirical approach (e.g., Fig.~\ref{fig-pa}) provides important insights that enable certain model solutions to be eliminated.  The empirical trends for $>30^{\rm d}$ Cepheids (Fig.~\ref{fig-pa}) should presently be interpreted cautiously owing to poor statistics, however, certain models predict that a turnover may occur near that boundary whereby metal-poor Cepheids exhibit the larger amplitudes.

\begin{figure*}[!t]
\begin{center}
\includegraphics[width=13cm]{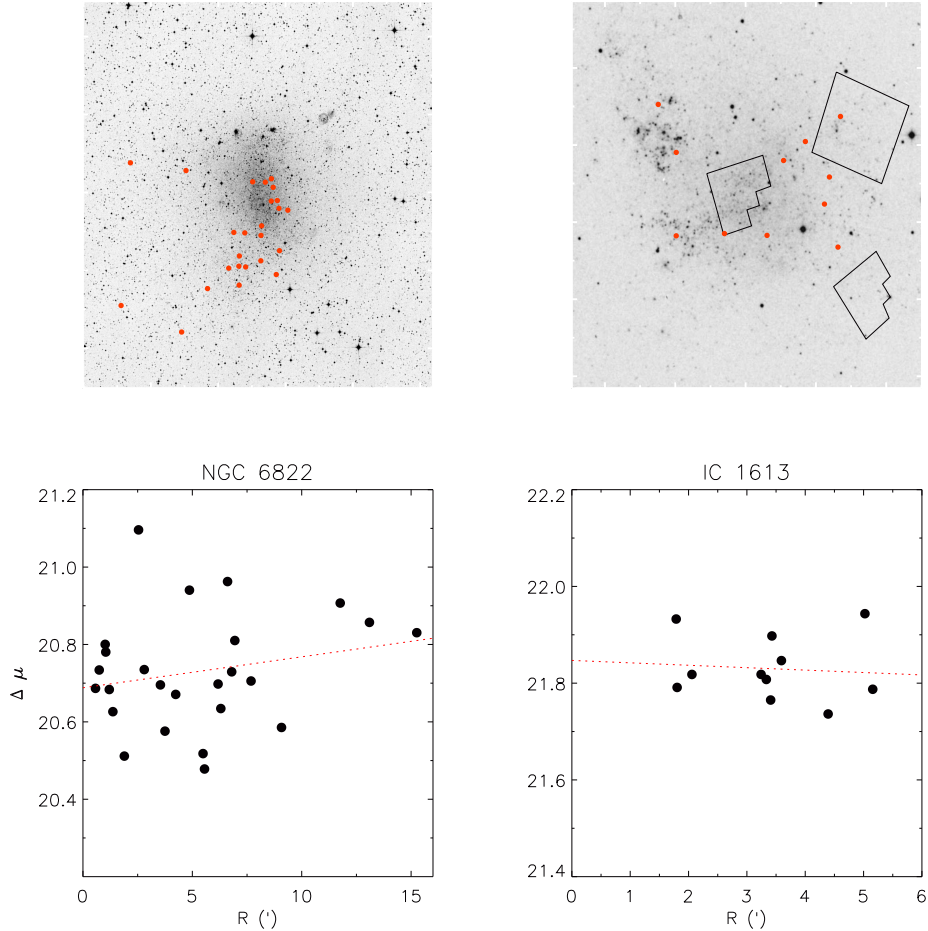} 
\caption{\small{Top panels, DSS images for NGC 6822 and IC 1613.  The red dots represent brighter Cepheids sampled by the Araucaria and OGLE surveys.  Approximate locations for the HST WFPC2 and ACS samplings are likewise shown.  A $\sim26^{\rm d}$ Cepheid residing away from the edge of the CCD detector exhibits both OGLE and ACS LCID photometry, and appears comparatively unblended ($\Sigma F_{*,V}/F_{Cep,V}<1$\%). Bottom panels, relative Wesenheit computed distances plotted as a function of the galactocentric distance.  The slopes of the best-fit trends are insignificant, yielding $0.008\pm0.008$ and $-0.005\pm0.019$ for NGC 6822 and IC 1613, respectively.  Blending does not appear to be the principal source behind the amplitude trends highlighted in Figs.~\ref{fig-pa} and \ref{fig-med}.}}
\label{fig-fov}
\end{center}
\end{figure*}

\subsection{{\rm \footnotesize BLENDING}}
\label{s-blending}
Setting a lower envelope to the period-amplitude diagram reveals that several stars in the Magellanic Clouds appear peculiar or blended, although the bulk of the sample appear typical.\footnote{$\sim10.5^{\rm d}$ Cepheids exhibit an amplitude minimum owing to the Hertzsprung progression \citep[see also][]{bo00b,ks09}, which may be metallicity dependent.}  Blending arising from contamination by neighboring stars in crowded extragalactic fields (or via binarity\footnote{\url{http://konkoly.hu/CEP/intro.html} , see also \citet{ev95,sz03}.}) can artificially dampen a Cepheid's amplitude.  A clear precedent exists which fuels such concerns. \citet{mo04} and \citet[][and references therein]{ma11b} asserted that photometric contamination compromises the Cepheid distance scale, since variables sampled near the cores of galaxies typically appear brighter than their counterparts in the outer (less dense and lower surface brightness) regions \citep[e.g., M101, M33,][]{ke98,ch12}. Blending has also propagated pernicious systematic effects into RR Lyrae distances, since variables found near the high-density cores of globular clusters can be spuriously brighter than their counterparts located near the cluster periphery \citep[][and references therein]{ma12,ma12b}. The photometric contamination affects other other globular cluster stars, complicates isochrone fitting \citep[][their Fig.~3]{ma12}, and introduces insidious degeneracies as the effect may mimic the signatures of differential reddening, (putative) colour-gradients, and multiple populations.  

The severity of blending is a function of numerous parameters, such as total seeing attributable to the observing system and environment, brightness of the Cepheid, star densities and distance to the target population, etc. The galaxy hosting the smaller-amplitude Cepheids is also the most distant (i.e., IC 1613). The Araucaria distance for IC 1613 \citep[$\sim720$ kpc,][]{pi06} places it well beyond the LMC, SMC, and NGC 6822, which have Araucaria established distances of $\sim52$ kpc, $62$ kpc, and $460$ kpc, respectively \citep{gi06,sz08,sz09}.  However, NGC 6822, IC 1613, and the SMC are comparatively uncrowded (Fig.~\ref{fig-fov}), particularly with respect to other Local Group galaxies such as M31 and M33.  NGC 6822 is also nearly an order of magnitude more distant than the SMC and exhibits marginally larger amplitudes, thereby hinting that blending is not causing the principal trend delineated in Figs.~\ref{fig-pa} and \ref{fig-med}.

A holistic approach relying on a suite of different evaluations can be used to identify blending.  In particular, a comparison between higher-resolution HST photometry and lower-resolution ground-based data is desirable \citep[e.g., M33,][]{pm11}.  The latter suffer from increased photometric contamination since ground-based seeing is typically an order of magnitude larger than HST seeing.  A single long-period Cepheid in IC 1613 exhibits published HST and OGLE photometry that were readily accessible to the authors. The variable displays a $\sim26^{\rm d}$ pulsation period, is located at J2000 01:01:57.4 +01:54:02.3, and is representative of longer-period Cepheids in IC 1613 (Fig.~\ref{fig-fov}).  A sizable fraction of the longer-period Cepheids sampled in IC 1613 (\& NGC 6822) occupy the galaxy's halo (Fig.~\ref{fig-fov}), where the stellar density and surface brightness are comparatively low.  There are 23 ACS-detected stars neighboring the Cepheid which fall well within the OGLE seeing, where the HST ACS photometry is from the LCID (Local Cosmology from Isolated Dwarfs) project \citep{be10,mo10}.  The sum of the $V$-band flux for those stars, divided by the mean Cepheid $V$-band flux ($\Sigma F_{*,V}/F_{Cep,V}$), yields a ratio less than $1\%$.  That indicates blending has a negligible impact on the amplitude of that Cepheid.   Several significantly fainter short-period Cepheids ($P<3.1^{\rm d}$) exhibit HST ACS and OGLE photometry, and $\sim15$\% of the flux measured for $\sim2-3^{\rm d}$ Cepheids appears to stem from blending.  The shortest-period Cepheid ($\sim10.45^{\rm d}$) used in Fig.~\ref{fig-pa} is $\sim1^{\rm m}.5$ brighter than those $2-3^{\rm d}$ Cepheids.  The median flux contributed by blending for the $2-3^{\rm d}$ sample reduces the amplitude of the $10.45^{\rm d}$ Cepheid by $0^{\rm m}.03$, which is insufficient to account for the offset observed in Fig.~\ref{fig-pa}.  Yet there exist blended long-period Cepheids in each galaxy investigated (Fig.~\ref{fig-pa}), and that includes the $14.33^{\rm d}$ Cepheid in IC 1613, which appears overly bright for its period and displays an amplitude well below the lower-bound approximated in Fig.~\ref{fig-pa}. 

Another pertinent test is to evaluate whether computed distance moduli for Cepheids in IC 1613 and NGC 6822 exhibit a galactocentric dependence (Fig.~\ref{fig-fov}).  Distance moduli calculated for classical Cepheids in M33 and M106 exhibit such a positional dependence \citep[][their Fig.~4]{ma09}. Cepheids in those galaxies appear brighter as the cluster core is approached, in concert with the increased star density and surface brightness. An inspection of the distance moduli estimated for brighter Cepheids in IC 1613 and NGC 6822 reveals no significant trend (Fig.~\ref{fig-fov}), where relative distances were computed via the Wesenheit function:
\begin{eqnarray}
\label{eqn-wvi}
\nonumber
W_{VI_c}-W_{VI_c,0}= \mu_0 \\
\nonumber
V-R_{VI_c}(V-I_c)-\alpha \log{P_0}-\beta=\mu_0 \\
V-R_{VI_c}(V-I_c)-\alpha \log{P_0}=\mu_0+\beta=\Delta \mu
\end{eqnarray}
$R_{VI_c}=2.55$ is the ratio of total to selective extinction \citep[][and references therein]{fo07}, $\alpha\sim-3.32$ is the slope of the Wesenheit function \citep{be07,so08,so10,ng09,ma11a}, and $\beta$ is chosen to be arbitrary (relative distance).

\begin{figure}[!t]
\begin{center}
\includegraphics[width=9cm]{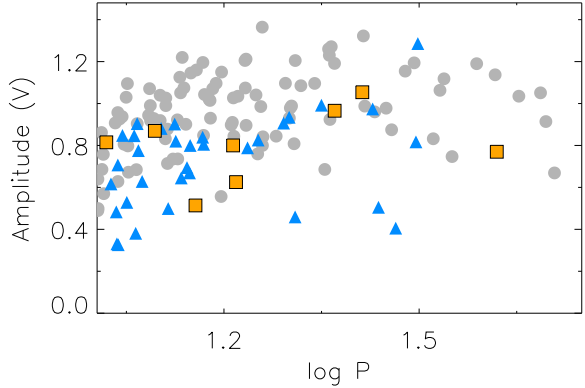} 
\caption{\small{Period-amplitude diagram for classical Cepheids in NGC 3109 (blue triangles), IC 1613 (orange squares), and the Milky Way (gray dots).  Cepheids in NGC 3109, like their very metal-poor counterparts in IC 1613 ($[{\rm Fe/H}]\sim-1$), typically exhibit smaller $V$-band amplitudes than metal-rich Galactic Cepheids.}}
\label{fig-ngc3109}
\end{center}
\end{figure}

\citet{ma10} note that $VI_c$ period-colour and period-Wesenheit relations can be employed, in tandem with other methods, to facilitate the identification of spurious photometry and blending. \citet{bo10} and \citet[][and references therein]{ma11b} argue that such functions are comparatively insensitive to variations in chemical composition.\footnote{For alternative interpretations and broader discussion see \citet{ro08}, \citet{fe11}, \citet{ma11b}, and \citet{mat12}.} It follows that negative mean reddenings inferred for Cepheid samples from $VI_c$ period-colour relations, or mean reddenings that are less than foreground estimates (IRAS dust maps), may be indicative of photometric problems \citep[Fig.~6 in][]{ma10}. Of similar note, significant deviations from the $VI_c$ Wesenheit slope of $\alpha\sim-3.32$ may also be suggestive of problematic photometry.  For example, ground-based photometry of Cepheids near the crowded core of M33 follow $\alpha\sim-2.8$ \citep[see also the pertinent discussion in][concerning the impact of blending on M33 Cepheid photometry]{pm11,ch12}.  The mean reddenings and Wesenheit slopes inferred from ground-based photometry for Cepheids in the Local Group galaxies analyzed here do not display such troubling trends \citep{ma10}.   

In sum, the principal amplitude offset shown in Figs.~\ref{fig-pa} and \ref{fig-med} likely results from variations in chemical composition, rather than blending.   Ultimately, a solid conclusion awaits the acquisition of high-resolution HST photometry for the extragalactic fields monitored using ground-based telescopes (Fig.~\ref{fig-fov}).

\subsection{{\rm \footnotesize NGC 3109}}
\label{s-ngc3109}
NGC 3109 is located at the periphery of the Local Group, and features a classical Cepheid distance of $d=1.30\pm0.02$ Mpc \citep{pi06b,so06}.\footnote{That Araucaria-based Cepheid distance agrees with a mean inferred from 18 published estimates tied to other methods: $d=1.33\pm 0.08(\sigma_{\bar{x}}) \pm 0.34 (\sigma )$ Mpc \citep[NED-D,][]{sm11}.}  The mean abundance of the Cepheid population can be estimated using the amplitude trends defined in \S \ref{s-analysis}, and the $V$-band observations acquired by \citet{pi06b}.  A period-amplitude diagram (Fig.~\ref{fig-ngc3109}) confirms that Cepheids in NGC 3109 typically display smaller $V$-band amplitudes than their metal-rich counterparts in the Milky Way, and are analogous to Cepheids in very metal-poor galaxies (SMC \& IC 1613).  Fig.~\ref{fig-pa} and the median amplitude ratio (Fig.~\ref{fig-med}) can be employed in concert to estimate the mean abundance of the Cepheid population, thereby mitigating the impact of high and low-amplitude outliers.  The resulting median amplitude ratio relative to the Milky Way sample is $0.81\pm0.03$, which together with the distribution displayed in Fig.~\ref{fig-ngc3109}, implies that Cepheids in NGC 3109 may potentially be more metal-poor than their SMC counterparts.  However, uncertainties tied to the present analysis prevent a definitive determination, particularly granted shorter-period ($P<19^{\rm d}$) Cepheids in NGC 3109 display increased scatter and blending appears more acute (Fig.~\ref{fig-ngc3109}).  Abundances inferred from HII regions and 8 B-type supergiants corroborate the aforementioned assertion that NGC 3109 Cepheids are very metal-poor like those in the SMC and IC 1613 \citep[][and references therein]{ev07}.  Thus the metallicity estimated from Cepheid pulsation matches that established by other means, to first order.

\section{{\rm \footnotesize CONCLUSION}}
A period-amplitude analysis was conducted of longer-period classical Cepheids in the Milky Way, LMC, NGC 6822, SMC, NGC 3109, and IC 1613. That sample spans a sizable abundance baseline, thereby enabling the impact of metallicity on Cepheid light amplitudes to be assessed.  On average, longer-period metal-rich classical Cepheids display larger $V$-band amplitudes than their counterparts in very metal-poor galaxies (Figs.~\ref{fig-pa}, \ref{fig-med}, \ref{fig-ngc3109}).  In particular, there is a readily discernible offset between relatively metal-rich (MW/LMC) and very metal-poor Cepheids (SMC/NGC 3109/IC 1613).  The resulting empirically deduced amplitude-metallicity dependence agrees with prior findings by \citet{vg78} and \citet{bi86}, and should be useful for constraining theoretical models, identifying blended (owing to binarity or crowding) and peculiar Cepheids, inferring approximate mean abundances for target Cepheid populations (\S \ref{s-ngc3109}) provided the sample is large enough \citep[see also][]{vg78,ta05}, ensuring consistent Cepheid/nebular abundances \citep{br10}, and improving lightcurve template fits for Cepheids with limited observational data.  The empirical results shown here lend support to models \citep[e.g.,][]{bo00} indicating that metal-rich $10-30^{\rm d}$ Cepheids could exhibit larger amplitudes than their metal-poor counterparts.  Further research assessing the impact of blending on Cepheid distances and parameters is desirable, and indeed, the ramifications are grave if the principal amplitude trend described here stems from blending.

\subsection*{{\rm \scriptsize ACKNOWLEDGEMENTS}}
\scriptsize{DM is grateful to the following individuals and consortia whose efforts or advice enabled the research: OGLE (A. Udalski, I. Soszy{\~n}ski), LCID (E. Bernard), L. Szabados, P. Klagyivik, R. Luck, V. Kovtychuk, G. Bono, C. Ngeow, H. Neilson, NED (I. Steer), M. Mottini, G. Benedict, A. van Genderen, CDS, arXiv, and NASA ADS. WG is grateful for support from the BASAL Centro de Astrofisica y Tecnologias Afines (CATA) PFB-06/2007.}

\end{document}